# Recently synthesized $(Zr_{1-x}Ti_x)_2AlC$ ($0 \leq x \leq 1$) solid solutions: Theoretical study of the effects of M mixing on physical properties

M. A. Ali[a*], M. M. Hossain[a], M. A. Hossain[b], M. T. Nasir[c], M. M. Uddin[a], M. Z. Hasan[d], A. K. M. A. Islam[d,e], S. H. Naqib[e**]

[a]Department of Physics, Chittagong University of Engineering and Technology (CUET), Chittagong-4349, Bangladesh.
[b]Department of Physics, Mawlana Bhashani Science and Technology University, Santosh, Tangail-1902, Bangladesh
[c]Department of Arts & Science, Bangladesh Army University of Science and Technology, Saidpur-5310, Nilphamari, Bangladesh.
[d]Department of Electrical and Electronic Engineering, International Islamic University Chittagong, Kumira, Chittagong -4318, Bangladesh.
[e]Department of Physics, University of Rajshahi, Rajshahi-6205, Bangladesh.

**Abstract:** The effects of M atomic species mixing on the structural, elastic, electronic, thermodynamic and charge transport properties of newly synthesized MAX phase $(Zr_{1-x}Ti_x)_2AlC$ ($0 \leq x \leq 1$) solid solutions have been studied for the first time by means of density functional theory (DFT) based first principles calculations. The lattice constants in good accord with the experimental results, are found to decrease with Ti content. The elastic constants, $C_{ij}$, and the other polycrystalline elastic moduli have been calculated. The elastic constants satisfy the mechanical stability conditions of these solid solutions. The constants $C_{11}$, $C_{33}$ and $C_{44}$ are found to increase with Ti contents up to $x = 0.67$, thereafter these decrease slightly. A reverse trend is followed by $C_{12}$ and $C_{13}$. The elastic moduli are also found to increase up to $x = 0.67$, beyond which these moduli go down slightly. Pugh's ratio and Poisson's ratio both confirm the brittleness of $(Zr_{1-x}Ti_x)_2AlC$. Different anisotropy factors revealed the anisotropic character of these solid solutions. A non-vanishing value of the electronic energy density of states (EDOS) at the Fermi level suggests that $(Zr_{1-x}Ti_x)_2AlC$ are metallic in nature. A mixture of covalent, ionic and metallic bonding has been indicated from the electronic structure with dominant covalent bonding due to hybridization of Zr-4$d$ states and C-2$p$ states. The variation of elastic stiffness and elastic parameters with $x$ is seen to be correlated with partial DOS (PDOS) and charge density distribution. The calculated Debye temperature and minimum thermal conductivity are found to increase with Ti contents, while melting temperature is the highest for $x = 0.67$. The solid solution with $x = 0.67$ shows improved mechanical and thermal properties compared to that of the two end members $Zr_2AlC$ and $Ti_2AlC$. The study of charge transport properties of $(Zr_{1-x}Ti_x)_2AlC$ reveals the metallic nature with saturated resistivity. The maximum power factor ($S^2\sigma/\tau = 11.1 \times 10^{10}$ $Wm^{-1}K^{-2}s^{-1}$) is obtained at 400 K for $(Zr_{1-x}Ti_x)_2AlC$.

Corresponding authors: *ashrafphy31@gmail.com; **salehnaqib@yahoo.com

## 1. Introduction

Since nearly two decades back [1], interest had been growing steadily on the $M_{n+1}AX_n$ phases (n = integer, M = early transition metal; A = group (13-16) element and X = C or N). It all started by the report on the remarkable properties of $Ti_3SiC_2$, discovered in the 1960s [2] in powder form. The intensive interest has grown on the MAX phases in the mid-1990s when the bulk form and the intrinsic properties of them were becoming well known. Consequently, the MAX phases have attracted considerable attention of research community as promising engineering materials due to their unique properties combining the merits of both metals and ceramics [3-11]. Like metals, these nano-laminates are electrically and thermally conductive, exceptionally damage tolerant, lightweight, and have good high-temperature mechanical properties, satisfactory resistance to oxidation and corrosion [3-5]. The key factor for the metallic and ceramic hybrid lies in their structure that consists of $M_{n+1}X_n$ sheets (for example, TiC) sandwiched in between one-atom thick A-layers (Si, Al, Ga, etc.) [3, 4]. Some phases also show self-healing characteristics [6, 7] and reversible deformation [8]. As of today, the number of synthesized ternary MAX phases reaches almost 80 individual compounds [12, 13], out of a possible 665 viable MAX phases ($M_{n+1}AX_n$, n =1–4) [14-26].

Concern regarding safety concepts for future fission reactors has received significant attention after the Fukushima's nuclear disaster. The concept of Accident Tolerant Fuels (ATF) is one that can resist the hostile atmosphere within a fission reactor for at least 10 hours in a Loss of Coolant Accident (LOCA). MAX phases can be chosen as cladding materials for use in ATF. The ternary $Zr_2AlC$ [27] phase is the most versatile materials among the MAX compounds. Zr is compatible with the zircaloy cladding, Al has a good resistance to corrosion and oxidation, while C limits nuclear transmutation. The $Ti_2AlC$ is currently considered as the most used MAX phase at high temperatures due to good mechanical properties and also good resistance to oxidation at high temperatures [28]. Therefore, solid solutions of the Zr-Ti-Al-C could be promising materials for using in the nuclear industry at high temperatures.

The prospect of partial substitution on the M, A, or X sites is an efficient way for forming solid solutions and have attracted substantial attention in recent years [26, 29-34] for the opportunity of tuning some of the useful properties of MAX phases [35-37]. However, attaining an exact a priori defined composition for a solid solution is challenging. Very recently, Tunca et al. [38] have synthesized $(Zr,Ti)_{n+1}AlC_n$ system where $(Zr_{1-x}Ti_x)_2AlC$ (211) solid solutions are reported. Moreover, both experimental synthesis and theoretical studies of physical properties of $(Zr,Ti)_3AlC_2$ have also been reported [39-41]. These studies motivate us to perform a first principles study of $(Zr,Ti)_2AlC$ solid solutions for the first time.

Therefore, We aim to investigate the effect of Ti substitution for Zr atoms in $Zr_2AlC$ on the structural, elastic, electronic, thermodynamic and charge transport properties of $(Zr_{1-x}Ti_x)_2AlC$ (211) solid solutions. It should be noted that we have considered only the synthesized composition of $(Zr_{1-x}Ti_x)_2AlC$ ($0 \leq x \leq 1$) solid solutions to avoid any question regarding phase stability.

The organization of rest of the paper is as follows. Section 2 describes the computational methodology. Results are presented and discussed in Section 3. Section 4 consists of the important conclusions of this study.

## 2. Calculation methods

To analyze the effects of Ti atom substitution for Zr in the $Zr_2AlC$ the density functional theory (DFT) [42, 43] was employed via the Cambridge sequential total energy package (CASTEP) code [44]. In this code, first-principles quantum mechanics calculations are implemented by the plane-wave pseudopotential method. The generalized gradient approximation (GGA) of the Perdew-Burke-Ernzerhof (PBF) [45] was adopted as the exchange and correlation terms. The electrostatic interaction between valence electron and ionic core was represented by the Vanderbilt-type ultra-soft pseudopotentials [46]. The cut-off energies of 500 eV were set for all calculations to ensure convergence. A k-point mesh of 10×10×2 according to the Monkhorst-Pack scheme [47] was used for integration over the first Brillouin zone. Density mixing was used to the electronic structure and Broyden Fletcher Goldfarb Shanno (BFGS) geometry optimization [48] was applied to optimize the atomic configurations. Periodic boundary conditions were used to calculate the total energies of each cell. Elastic constants were calculated by the usual 'stress-strain' method.

The transport properties of $Zr_{1-x}Ti_x)_2AlC$ (x = 0, 0.5, 1) have also been studied by the combined use of full potential linearized augmented plane-wave (FP-LAPW) method implemented in WIEN2k code and BoltzTraP code [49, 50]. The GGA within the PBE scheme was used [45]. A plane wave cut-off of kinetic energy $R_{MT}K_{max}$ = 7.0 was selected by convergence tests. A 10000 k-points in the irreducible representations were used for self-consistent calculations. The muffin-tin radii for Zr, Ti, Al and C were fixed to 2.28, 2.28, 2.5 and 1.86 Bohr radii, respectively. BoltzTraP program was used to calculate the electronic transport properties, which solves the semiclassical Boltzmann transport equation using the rigid-band approach [51, 52]. The relaxation time $\tau$ was taken to be a constant. The electronic conductivity and the electronic part of thermal conductivity were calculated with respect to $\tau$, whereas the Seebeck coefficient was independent of $\tau$.

## 3. Results and discussion
### 3.1. *Structural properties*

$(Zr_{1-x}Ti_x)_2AlC$ crystallize in the hexagonal structure with space group $P6_3/mmc$, like other MAX phases [12]. The unit cell of $Zr_2AlC$ with two formula units is shown in Fig.1(a). The positions of atoms in $Zr_2AlC$ are as follows: C atoms are placed at the positions (0, 0, 0), the Al atoms are at (1/3, 2/3, 3/4) and the Zr atoms are at (1/3, 2/3, $z_M$). The Ti atoms are substituted for Zr at (1/3, 2/3, $z_M$). The equilibrium crystal structure of this compound is obtained by minimizing the total energy. Optimum structural parameters of pure ($Zr_2AlC$) and $(Zr_{1-x}Ti_x)_2AlC$ solid solutions are presented in Table 1. The results are in good agreement with their corresponding experimental results. Fig. 1(b) shows the lattice constant '*a*' and '*c*' of Ti substituted $(Zr_{1-x}Ti_x)_2AlC$ compounds where the effect of Ti substitution is clearly observed. The Vegard's law [53] is followed by each composition. The lattice constant of the substituted compounds is found to decrease with Ti content due to the ionic size differences between Zr (0.89 Å) and Ti (0.74 Å). Since the Zr atoms are partially replaced by Ti atoms, it is expected to cause contraction of the unit cell. The similar trend for both experimental and calculated lattice constants is observed. The hexagonal ratio (*c/a*) showed in Fig. 1(b) increases as the Ti content *x* increases, which reveals that the values of *c* decrease less significantly than that of *a*. It also means that the compressibility along c-axis decreases with increasing of Ti content *x*.

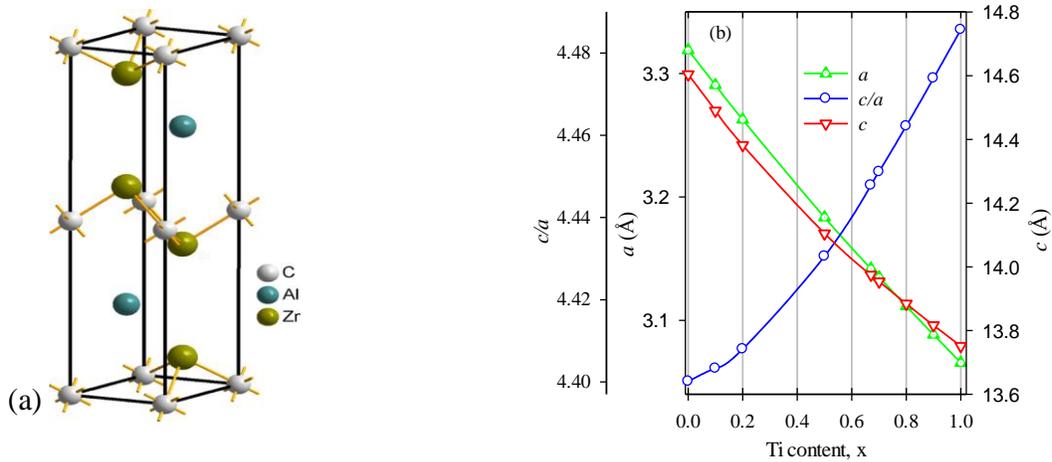

Fig. 1. (a) Crystal structure of the $Zr_2AlC$ compound and (b) lattice constants of $(Zr_{1-x}Ti_x)_2AlC$ as a function of Ti contents (*x*).

**Table 1**
Calculated lattice parameters (*a* and *c*), *c/a* ratio, and unit cell volumes (*V*) for the $(Zr_{1-x}Ti_x)_2AlC$ ($0 \leq x \leq 1$) solid solutions, together with other available results.

| x | a (Å) | c (Å) | c/a | V (Å³) | Ref. |
|---|---|---|---|---|---|
| 0.00 | 3.319 | 14.6045 | 4.40 | 139.300 | This |
|  | 3.3230 | 14.5700 | 4.38 | 139.328 | [38] |
|  | 3.319 | 14.6045 | 4.40 | 139.300 | [54] |
| 0.10 | 3.2906 | 14.4899 | 4.40 | 135.881 | This |
|  | 3.3140 | 14.5070 | 4.37 | 137.975 | [38] |
| 0.20 | 3.2627 | 14.3819 | 4.40 | 132.583 | This |
|  | 3.2656 | 14.3630 | 4.39 | 132.644 | [38] |
| 0.50 | 3.1834 | 14.1047 | 4.43 | 123.791 | This |
|  | 3.1813 | 14.0204 | 4.40 | 122.881 | [38] |
| 0.67 | 3.142 | 13.9751 | 4.44 | 119.477 | This |
|  | 3.1627 | 13.9534 | 4.41 | 120.868 | [38] |
| 0.70 | 3.1349 | 13.9538 | 4.454 | 118.756 | This |
|  | 3.1583 | 13.9307 | .41 | 120.336 | [38] |
| 0.80 | 3.1115 | 13.8843 | 4.46 | 116.407 | This |
|  | 3.1314 | 13.8381 | 4.41 | 117.509 | [38] |
| 0.90 | 3.0883 | 13.8166 | 4.47 | 114.119 | This |
|  | 3.0919 | 13.7500 | 4.44 | 113.833 | [38] |
| 1.00 | 3.0655 | 13.7508 | 4.48 | 111.904 | This |
|  | 3.0596 | 13.6470 | 4.46 | 110.633 | [38] |
|  | 3.066 | 13.694 |  | 111.49 | [55] |
|  | 3.058 | 13.642 |  | 110.48 | [56] |

### 3.2 *Mechanical properties*

The study of mechanical properties such as elastic moduli, elastic anisotropy and ductile or brittle behavior are of critical importance for the wear resistance of surface coating materials. The elastic constants are calculated by computing the resulting stress which is created due to applying a set of finite homogeneous deformation within the CASTEP code from first principles method [57]. Different elastic constants and elastic moduli of $(Zr_{1-x}Ti_x)_2AlC$ ($0 \leq x \leq 1$) have been calculated and presented in Table 2. Five independent elastic constants ($C_{ij}$) are shown in Table 2. These elastic constants completely satisfy the requirement of mechanical stability in a hexagonal crystal following Born [58] restrictions on the elastic constants, namely: $C_{11} > 0$, $C_{11}-C_{12} > 0$, $C_{44} > 0$, $(C_{11}+C_{12})C_{33} - 2C_{13} > 0$. Therefore, the MAX $(Zr_{1-x}Ti_x)_2AlC$ ($0 \leq x \leq 1$) solid solutions under consideration are mechanically stable.

The elastic tensors can also give information about the bonding nature of solids. The elastic constant $C_{11}$ gives the elasticity in length and measures the elastic stiffness of solids with respect to the (100) ⟨100⟩ uniaxial strain. The elastic constant $C_{33}$ corresponds to the uniaxial deformation along the ⟨001⟩ direction. From Table 2, it is evident that $C_{11} > C_{33}$ for all compositions indicating stronger bonding along the [100] direction than that of the [001] direction and revealed that it is more difficult to compress the solid solutions along the crystallographic *a*-axis, than along the *c*-axis. In fact, $C_{11} > C_{33}$ is the general trend in MAX phase carbides and reveals the anisotropic nature of the compound. The elastic tensors $C_{12}$ and $C_{44}$ are associated with the elasticity in shape. Particularly, $C_{12}$ corresponds to the pure shear stress in (110) plane along ⟨100⟩ direction, whereas $C_{44}$ signifies the shear stress in (010) plane in the ⟨001⟩ direction. The elastic constant $C_{13}$ is found to be almost equal to $C_{12}$. These two constants combine a functional stress component in the crystallographic $a$ direction in the presence of a uniaxial strain along the crystallographic $b$ and $c$ axes. The variation of $C_{ij}$ with Ti content is shown in Fig. 2 (a) and it is found that these constants increase rapidly up to $x = 0.67$, thereafter these become almost constant.

**Table 2**
The elastic constants $C_{ij}$ (GPa), bulk modulus, $B$ (GPa), shear modulus, $G$ (GPa), Young's modulus, $Y$ (GPa), Pugh ratio, $G/B$, Vickers hardness, $H_v$ (GPa), Poisson ratio, $v$ and Cauchy pressure for different Ti contents ($x$).

| $x$ | $C_{11}$ | $C_{12}$ | $C_{13}$ | $C_{33}$ | $C_{44}$ | $B$ | $G$ | $Y$ | $G/B$ | $H_v$ | $v$ | Cauchy Pressure | Ref. |
|---|---|---|---|---|---|---|---|---|---|---|---|---|---|
| 0.00 | 258 | 67 | 63 | 221 | 91 | 125 | 92 | 222 | 0.74 | 16.7 | 0.20 | -24 | This |
|      | 261 | 63 | 63 | 224 | 87 | 125 | 92 | 221 | 0.74 | 16.7 | 0.20 | -24 | [59] |
| 0.10 | 268 | 66 | 63 | 234 | 97 | 128 | 97 | 232 | 0.76 | 18.0 | 0.20 | -31 | This |
| 0.20 | 278 | 64 | 63 | 243 | 101 | 131 | 102 | 243 | 0.78 | 19.3 | 0.19 | -37 | This |
| 0.50 | 295 | 63 | 63 | 262 | 110 | 136 | 111 | 262 | 0.82 | 21.8 | 0.18 | -47 | This |
| 0.67 | 300 | 62 | 62 | 266 | 113 | 137 | 114 | 268 | 0.83 | 22.8 | 0.17 | -51 | This |
| 0.70 | 298 | 63 | 62 | 266 | 112 | 137 | 113 | 266 | 0.82 | 22.4 | 0.18 | -49 | This |
| 0.80 | 298 | 63 | 62 | 266 | 112 | 137 | 113 | 266 | 0.82 | 22.4 | 0.18 | -49 | This |
| 0.90 | 298 | 64 | 61 | 265 | 111 | 137 | 113 | 266 | 0.82 | 22.4 | 0.18 | -47 | This |
| 1.00 | 298 | 64 | 63 | 270 | 111 | 138 | 113 | 266 | 0.82 | 22.1 | 0.18 | -47 | This |
|      | 302 | 62 | 58 | 270 | 109 | 137 | 114 | 267 | 0.83 | 22.8 | 0.17 | -47 | [59] |

The bulk and shear moduli ($B$, $G$) are calculated from the single crystal zero-pressure elastic constants through the Voigt-Reuss-Hill formula [60, 61]. In addition, the Young's modulus ($Y$) and Poisson's ratio ($v$) can also be obtained using well-known relationships [62, 63]. All these calculated elastic parameters are presented in Table 2. The bulk modulus is an important property, used to express the average bond strength of constituent atoms for a given solid [64]. The calculated value of $B$ (128 GPa) indicates moderately strong average bonding strength of the atoms involved in $(Zr_{1-x}Ti_x)_2AlC$. The bond strength of atoms provides the required resistance to volume change under external pressure. On the other hand, deformation (i.e., shape change) in a solid crucially depends on its shear modulus $G$. Higher the value of $G$, the more rigid the material is. Although, these two parameters do not measure the hardness of the materials directly, Vickers hardness can be estimated using Chen's formula [65], given by: $H_v = 2(k^2G)^{0.585} - 3$, where, $k = G/B$. Although, it is said that a direct relation between the shear modulus and hardness is relatively greater than that with the bulk modulus [66], but the Pugh's ratio ($k = G/B$) indicate both these moduli are important. In order to discuss the effect of Ti atoms on the mechanical hardness, we have used Chen's formula to calculate the hardness as shown in Table 2. The hardness is found to increase with Ti contents up to $x = 0.67$, then slightly decrease with almost constant value (~ 22.4 GPa) up to $x = 1.0$. Note here that the value of hardness for solid solution with x= 0.67 was enhanced by 36.53% compared to $Zr_2AlC$, indicating the much improved mechanical properties of the solid solutions.

The value of $Y$ (219 GPa) indicates that the $Zr_2AlC$ is moderately stiff. In addition, the value of $Y$ increases with the increase of Ti contents, reaching an end value of 267 GPa for $Ti_2AlC$.

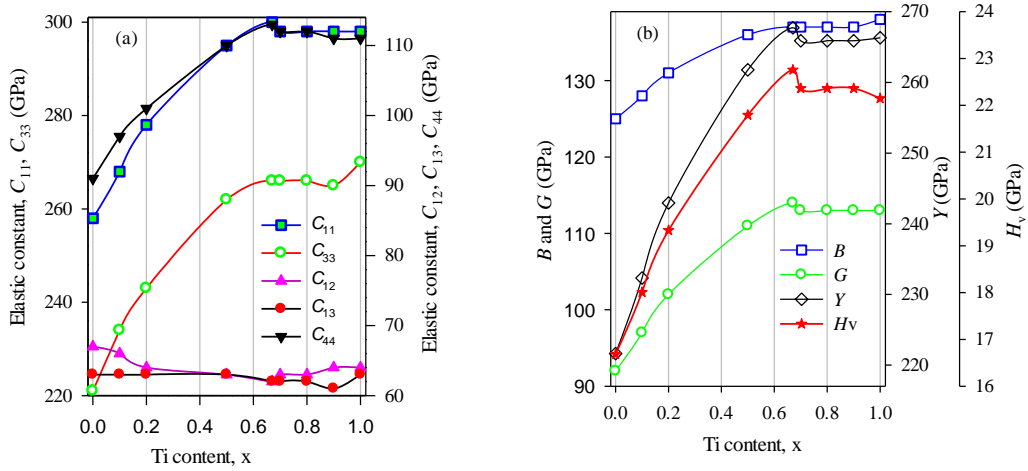

Fig. 2. Variation of (a) elastic constants, $C_{ij}$ and (b) elastic moduli and hardness with Ti contents ($x$).

A material is said to be brittle if the Pugh ratio G/B [67] is greater than 0.57; otherwise, the material is ductile. As is evident from Table 2, the values of Pugh ratio lies between 0.72-0.82. Therefore, the $(Zr_{1-x}Ti_x)_2AlC$ MAX phases are brittle materials which is a common trend in MAX phase compounds. In addition to Pugh ratio, Frantsevich's [68] also proposed a critical value of Poisson's ratio ($v \sim 0.26$) to separate the brittle and ductile nature of solids. Our calculated values lie within the range 0.18 - 0.20 for $(Zr_{1-x}Ti_x)_2AlC$, again demonstrating the brittleness of the materials under study.

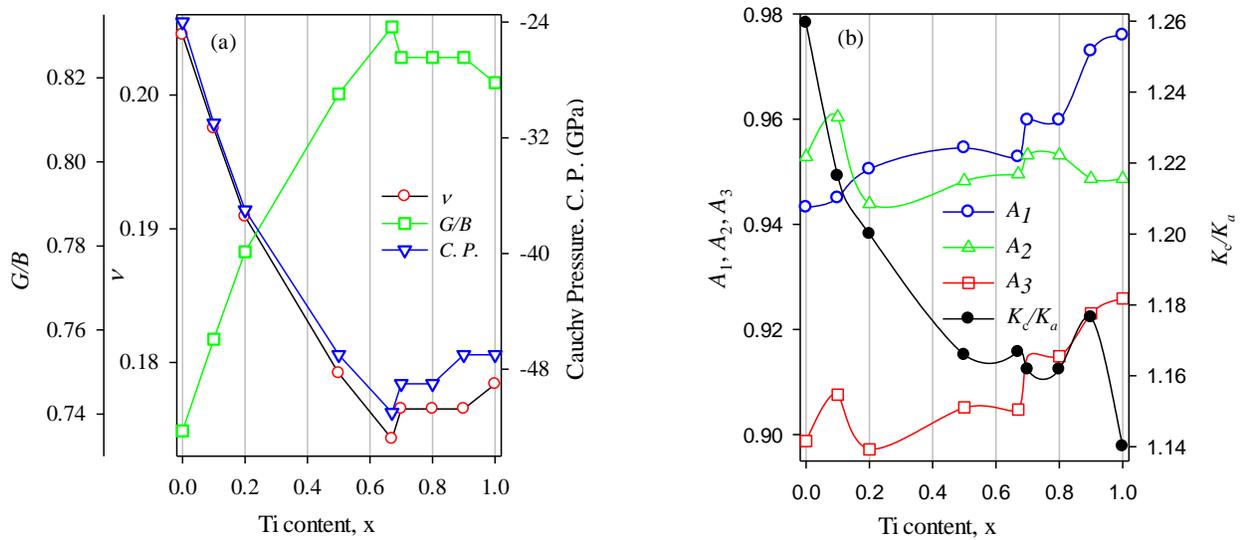

Fig. 3. Variation of (a) Pugh ratio, Poisson's ratio, Cauchy pressure and (b) different anisotropic constants with Ti contents ($x$).

Important information regarding the ionic/covalent bonding of the materials can also be obtained from the value of $v$. The value of $v$ is typically small ($v = 0.1$) for covalent materials and while the value of $v$ is around 0.25 [69] for the ionic materials. The calculated values of $v$ for the $(Zr_{1-x}Ti_x)_2AlC$ solid solutions lie within this border line of covalent/ionic materials indicating existence of mixture of covalent and ionic bonding in these compounds. The value Poisson's ratio is related to the nature of inter-atomic bonding forces. The range of $v$ for central-force solids are 0.25 to 0.5 [70]. Since the calculated Poisson's ratios do not belong into this range, non-central inter-atomic forces dominate in these solid solutions. The difference, $(C_{12} - C_{44})$, known as the Cauchy pressure, also gives information regarding the type of bonding in solids. The value of Cauchy pressure is positive for a crystal in which metallic bonding dominate. In case of a negative Cauchy pressure, the material is characterized by directional covalent bonding [71]. The negative value of Cauchy pressure for all compositions demonstrates the dominating nature of directional covalent bonding and the brittle nature as well.

Many physical processes are influenced by elastic anisotropy such as development of plastic deformation in crystals, enhanced charged defect mobility, micro-scale cracking in ceramics, phase transformations, mechanical yield points, elastic instability, internal friction, etc. [72]. The elastic anisotropy can be described in several ways for hexagonal crystal system. The degree of anisotropy in the bonding between atoms in different planes is described by the shear anisotropic factors which are obtained from $A_1 = \frac{1/6(C_{11}+C_{12}+2C_{33}-4C_{13})}{C_{44}}$, $A_2 = \frac{2C_{44}}{C_{11}-C_{12}}$, $A_3 = A_1 . A_2 = \frac{1/3(C_{11}+C_{12}+2C_{33}-4C_{13})}{C_{11}-C_{12}}$ [73] for {100} planes between the ⟨011⟩ and ⟨010⟩ directions, for {010} shear planes between the ⟨101⟩ and ⟨001⟩ directions and for {001} shear planes between the ⟨110⟩ and ⟨010⟩ directions, respectively. The calculated $A_i$'s in Table 3 show that the solid solutions are anisotropic (for isotropy $A_i = 1$) in three shear planes. The different calculated values of $A_1$, $A_2$ and $A_3$ also exhibit the strong directional dependence of shear modulus. The elastic anisotropy of $(Zr_{1-x}Ti_x)_2AlC$ solid solutions can be characterized by the values of the bulk modulus along the $a$-axis ($B_a$) and $c$-axis ($B_c$), and the expressions are as follows [74], $B_a = a\frac{dP}{da} = \frac{\Lambda}{2+\alpha}$, $B_c = c\frac{dP}{dc} = \frac{B_a}{\alpha}$, where $\Lambda = 2(C_{11} + C_{12}) + 4C_{13}\alpha + C_{33}\alpha^2$ and $\alpha = \frac{(C_{11}+C_{12})-2C_{13}}{C_{33}+C_{13}}$. The calculated values for $B_a$ and $B_c$ (Table 3) show that $B_c > B_c$ indicating that the compression along $a$-axis is more difficult than $c$-axis for all solid solutions considered here.

**Table 3**

Anisotropic factors, $A_1$, $A_2$, $A_3$, $k_c/k_a$, $B_a$, $B_c$, percentage anisotropy factors $A_G$ and $A_B$, and universal anisotropic index $A^U$ of $(Zr_{1-x}Ti_x)_2AlC$ solid solutions.

| Ti contents, x | $A_1$ | $A_2$ | $A_3$ | $k_c/k_a$ | $B_a$ | $B_c$ | $A_B$ | $A_G$ | $A^U$ |
|---|---|---|---|---|---|---|---|---|---|
| 0.00 | 0.94 | 0.95 | 0.90 | 1.26 | 404 | 321 | 0.0028 | 0.0010 | 0.0165 |
| 0.10 | 0.95 | 0.96 | 0.91 | 1.22 | 410 | 337 | 0.0020 | 0.0010 | 0.0144 |
| 0.20 | 0.95 | 0.94 | 0.90 | 1.20 | 417 | 348 | 0.0019 | 0.0011 | 0.0150 |
| 0.50 | 0.95 | 0.95 | 0.91 | 1.17 | 431 | 370 | 0.0014 | 0.0010 | 0.0132 |
| 0.67 | 0.95 | 0.95 | 0.90 | 1.17 | 434 | 372 | 0.0014 | 0.0010 | 0.0130 |
| 0.70 | 0.96 | 0.95 | 0.91 | 1.16 | 433 | 372 | 0.0013 | 0.0004 | 0.0074 |
| 0.80 | 0.96 | 0.95 | 0.91 | 1.16 | 433 | 372 | 0.0014 | 0.0007 | 0.0103 |
| 0.90 | 0.97 | 0.95 | 0.92 | 1.18 | 433 | 368 | 0.0017 | 0.0008 | 0.0122 |
| 1.00 | 0.98 | 0.95 | 0.93 | 1.14 | 433 | 380 | 0.0011 | 0.0007 | 0.0099 |

The elastic anisotropy in polycrystalline solids can also be calculated with the percentage anisotropy in compressibility and shear [75]. These two factors can be expressed as: $A_B = \frac{B_V - B_R}{B_V + B_R} \times 100\%$ and $A_G = \frac{G_V - G_R}{G_V + G_R} \times 100\%$, where the subscripts, $V$ and $R$, correspond to the Voigt and Reuss bounds, respectively. $B$ and $G$ are the bulk and shear moduli. Table 3 shows that the anisotropy is greater in compressibility than in shear for all compositions.

Another important measure of anisotropy, known as the universal anisotropic index $A^U$, is defined as: $A^U = 5\frac{G_V}{G_R} + \frac{B_V}{B_R} - 6 \geq 0$, where $A^U = 0$ is for isotropic materials and the departure from zero defines the extent of anisotropy [76]. The value of $A^U$ for the $(Zr_{1-x}Ti_x)_2AlC$ solid solutions are very close to zero. However, overall anisotropic factors (Table 3) reveal that the solid solutions under consideration are anisotropic.

### 3.3 Electronic properties

We now investigate the electronic band structure and DOS (TODS and PDOS) to understand how the electronic properties would differ in $(Zr_{1-x}Ti_x)_2AlC$ solid solutions as $x$ is varied. It also provides us with important information regarding the natures of chemical bonding and structural features. The calculated band structure and DOS of $Zr_2AlC$ are displayed in Figs. 4 (a and b). Considerable overlapping of valence and conduction band and the non-vanishing value of DOS at the Fermi level make the metallic character of the $Zr_2AlC$ quite clear. Al and C atoms only contribute weakly to the TDOS around $E_F$ and therefore are not significantly associated with the metal-like conduction. Instead, the Zr-$d$ electrons play the dominant role in the electronic conduction properties, with the Zr-$d$ PDOS making by far the largest contribution to the TDOS around $E_F$. The valence band can be decomposed into three well separated regions: (i) a deeply bound lowest valence band (LVB) between -11.0 eV to -9.0 eV energy range (ii) the mid valence band (MVB) between -7.0 eV to -2.0 eV and (ii) upper valence band (UVB) between -2.0 eV to 0.0 eV. The LVB, peaked below −11.0 eV, is mainly composed of the C-$2s$ states. The MVB is composed of C-$2p$ states and Zr-$4d$ states and the peak correspond to the Zr $4d$-C $2p$ bonding states. The Zr-C ($pd$) bonding is primarily covalent, because the states of Zr and C atoms are strongly mixed together. Strong $pd$ covalent bonding states dominate the higher energy range. The states just below the Fermi level contain relatively weaker $pd$ covalent bonding between Zr and Al. Moreover, states near and above the Fermi level are attributed to metal-to-metal $dd$ interactions and anti-bonding states. In order to understand the effect of Ti substitution, we have also calculated the TDOS and PDOS for substituted solid solution $(Zr_{0.5}Ti_{0.5})_2AlC$, which is shown in Fig. 4 (c). Quite similar overall DOS profiles are obtained for all solid solutions. Some distinctive differences can be traced to the features of the position of the peaks. Fig. 4 (d) shows the position of the peaks in TDOS for different compositions where two vertical solid green lines are used to refer the positions of these peaks. It is clearly observed that the position of the peaks at the UVB derived from hybridization between Zr/Ti-$d$ and C-$p$ states and Zr/Ti-$d$ and Al-$p$ states slightly shift towards the lower energy up to $x = 0.67$, giving a better energy overlap between different states and hence strengthening the covalent bonding between them. It can also be observed that the $d$ and $p$ states slightly shift towards the higher energy states after $x = 0.67$ and hence weakening the covalent bonding between Zr/Ti-$4d$ states and C-$p$ states and Zr/Ti-$d$ states & Al-$p$ states. Therefore, the most strong $pd$ covalent bonding is observed for $x = 0.67$ which results the hardest solids solution among the studied compositions.

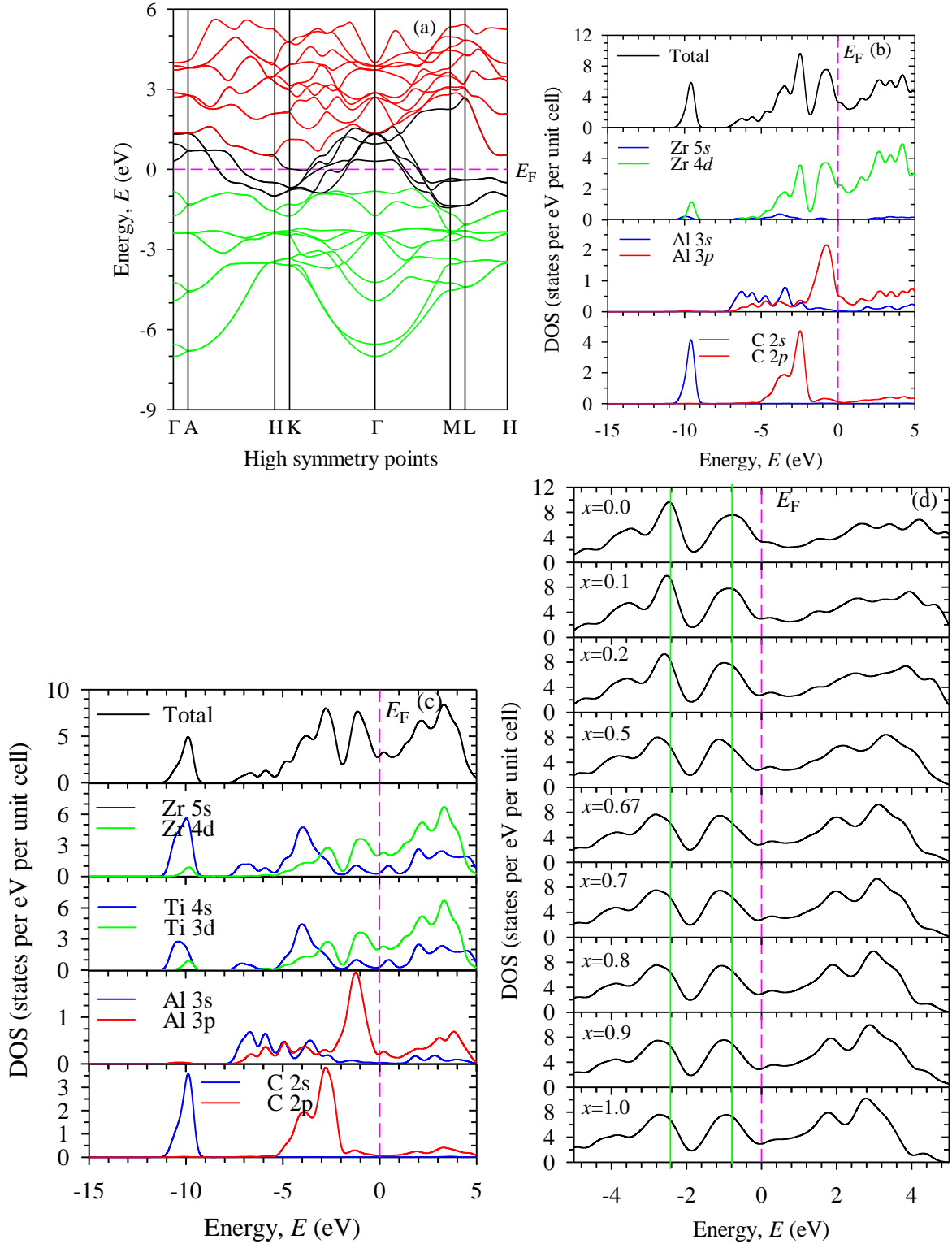

Fig. 4 (a) Electronic band structure and (b) density of states of $Zr_2AlC$, (c) density of states of $(Zr_{0.5}Ti_{0.5})_2AlC$, (d) total density of states for different Ti contents ($x$).

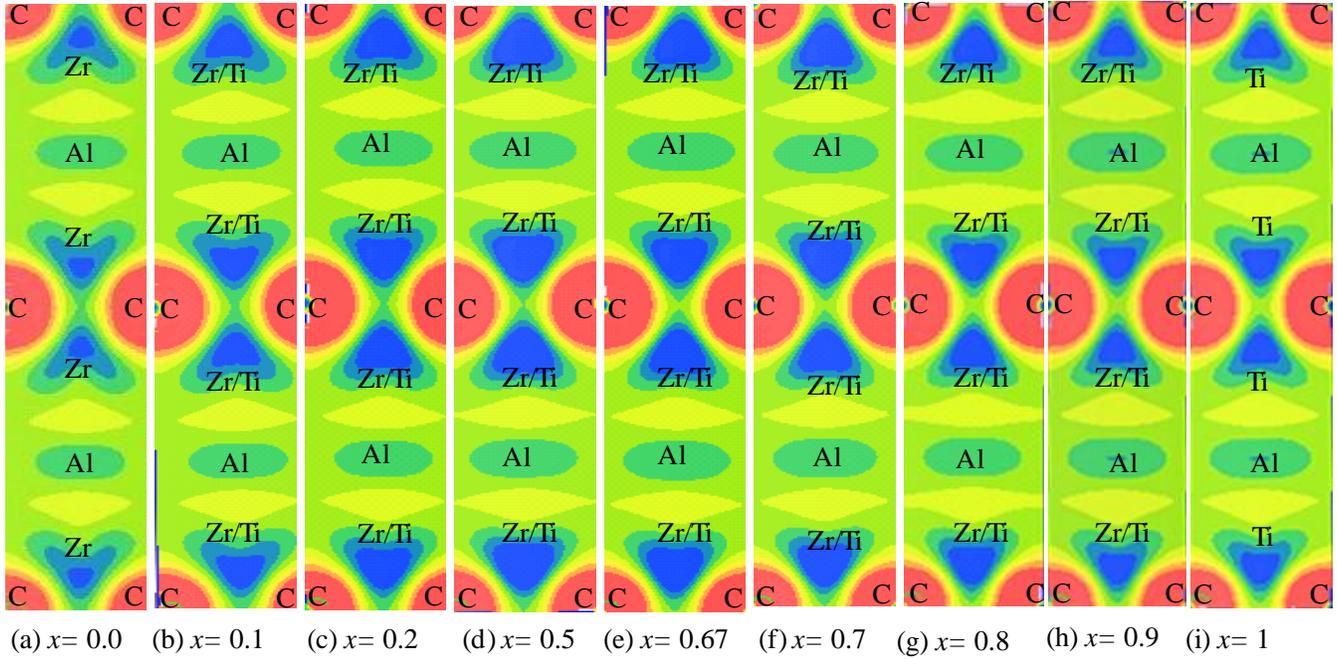

(a) $x= 0.0$ (b) $x= 0.1$ (c) $x= 0.2$ (d) $x= 0.5$ (e) $x= 0.67$ (f) $x= 0.7$ (g) $x= 0.8$ (h) $x= 0.9$ (i) $x= 1$

**Fig. 5** The electronic charge density for different Ti contents ($x$).

*Charge density*

In order to describe the nature of chemical bonding further in $(Zr_{1-x}Ti_x)_2AlC$, electron charge density mapping (in the units of e/Å$^3$) along (101) crystallographic plane is calculated as shown in Fig. 5. The feature of this mapping is the exhibition of the electron densities due to different chemical bonds. The accumulation and depletion of electronic charges can be understood from the regions of positive or negative electrons density, respectively. The accumulation of charges between two atoms exhibits the covalent bonds where as the balancing of positive or negative charge at the atomic position indicates the ionic bonding [77]. From Fig. 5, it is evident that most charge accumulated regions are the positions of C atoms and Zr atoms. As a result, the strong covalent bonding between Zr and C atoms is formed. There is also another covalent bonding between Zr and Al atoms which is comparatively weaker in strength. It is also observed that the density of accumulating Zr electrons increases with increasing Ti contents up to $x = 0.67$ and thereafter it decreases slightly. Therefore, the covalent bonding between Zr/Ti and C atoms should be strongest for $x = 0.67$ which is consistent with elastic constants, hardness as well as DOS results. From the analysis of the electronic properties, it is worth noting that the covalent bonding of Zr/Ti-C is much more stronger than that of Zr/Ti-Al for all the $(Zr_{1-x}Ti_x)_2AlC$ compounds which is commonly seen in the MAX phases due to stacks of 'hard' M-X bond and 'soft' M-A bond along the $c$ direction [4]. According to the correlation between the shear modulus, bulk modulus and hardness, the $(Zr_{0.33}Ti_{0.67})_2AlC$ compound is expected to possess higher hardness than most other MAX compounds.

### 3.4 *Thermodynamic properties*

We have investigated the important thermodynamic properties such as Debye temperature, Grüneisen parameter, melting temperature of the $(Zr_{1-x}Ti_x)_2AlC$ solid solutions.

The Debye temperature, $\Theta_D$ is an essential parameter of solids used to describe some physical process related to phonons, lattice vibration enthalpy, thermal conductivity, melting point, specific heat. The $\Theta_D$ can be calculated by a simple method using average sound velocity proposed by Anderson [78]. The calculation details can be found elsewhere [79-81]. The different elastic wave (sound wave) velocities and Debye temperature $\Theta_D$ of the $(Zr_{1-x}Ti_x)_2AlC$ solid solutions have been calculated and listed in Table 4 and their variation with Ti contents are shown in Fig. 6. It is well-known that the sound velocities are dependent on the bulk modulus, shear modulus and density of a material. It means that if a material has large values of the elastic moduli, correspondingly it will show a higher sound velocity. Due to replacement of Zr atoms with comparatively lighter element Ti atoms, the density of solid solutions decreases with Ti. Therefore, although the $B$ and $G$ of the solid solutions are highest for $x = 0.67$, but due to lower density the longitudinal, transverse and mean sound velocities as well as the $\Theta_D$ are found to gradually increase with Ti contents.

**Table 4**

Calculated density, longitudinal, transverse and average sound velocities ($v_l$, $v_t$, and $v_m$), Debye temperature $\Theta_D$, minimum thermal conductivity $K_{min}$, Grüneisen parameter $\gamma$ and melting temperature $T_m$ of $(Zr_{1-x}Ti_x)_2AlC$ solid solutions.

| Ti x | $\rho$ (gm/cm$^3$) | $v_l$ (m/s) | $v_t$ (m/s) | $v_m$ (m/s) | $\Theta_D$ (K) | $K_{min}$ (W/(mK)) | $\gamma$ | $T_m$ (K) |
|---|---|---|---|---|---|---|---|---|
| 0.00 | 5.278 | 6849.13 | 4174.41 | 4610.71 | 529.70 | 1.44 | 1.30 | 1459 |
| 0.10 | 5.200 | 7034.42 | 4318.83 | 4766.64 | 552.16 | 1.51 | 1.27 | 1509 |
| 0.20 | 5.112 | 7226.62 | 4466.63 | 4926.31 | 575.35 | 1.58 | 1.25 | 1552 |
| 0.50 | 4.777 | 7709.88 | 4820.03 | 5309.58 | 634.46 | 1.79 | 1.20 | 1632 |
| 0.67 | 4.540 | 7978.05 | 5010.72 | 5516.85 | 667.07 | 1.90 | 1.19 | 1653 |
| 0.70 | 4.495 | 7999.54 | 5013.71 | 5521.42 | 668.97 | 1.91 | 1.20 | 1647 |
| 0.80 | 4.338 | 8142.71 | 5103.44 | 5620.24 | 685.50 | 1.97 | 1.20 | 1647 |
| 0.90 | 4.173 | 8302.46 | 5203.57 | 5730.50 | 703.59 | 2.04 | 1.20 | 1645 |
| 1.00 | 3.998 | 8496.71 | 5316.08 | 5855.52 | 723.65 | 2.11 | 1.20 | 1653 |

The MAX phases materials are potential candidates for applications at high temperatures. When temperature reaches to very high values the lattice thermal conductivity converges to its minimum value $k_{min}$ which is independent of temperature due to complete uncoupling of phonons during transfer of heat energy between neighboring atoms [82]. At this point, the average inter-atomic distance can be considered as the phonon mean free path. Consequently, an "equivalent atom" can be considered as a replacement of different atoms in a molecule where the "equivalent atom" should have average atomic mass of M/n (n = number atoms in the unit cell). As a result, a single "equivalent atom" in the cell does not have any optical modes and the minimum thermal conductivity $k_{min}$ is obtained by the equation [83] $k_{min} = k_B v_m \left(\frac{M}{n\rho N_A}\right)^{-2/3}$, where $k_B$, $v_m$, $N_A$ and $\rho$ are Boltzmann constant, average phonon velocity, Avogadro's number and density of crystal, respectively.

The Grüneisen parameter $\gamma$ gives a measure of the anharmonic effect in crystalline solids. For instance, the temperature dependence of phonon frequencies as well as line-widths and the thermal expansion effects. The Grüneisen parameter [84], is defined by $\gamma(\omega_n) = \frac{d\ln(\omega_n)}{d(\ln\varphi)}$ with $\omega_n$ and $\varphi$ are the frequency and packing fraction of the crystal, respectively. We have calculated the Grüneisen parameter by a simple expression that relates the Grüneisen parameter $\gamma$ to the Poisson ratio, $\upsilon$ i.e., $\gamma = \frac{3}{2}\frac{(1+\nu)}{(2-3\nu)}$ [84].

In addition, the melting temperature, $T_m$, is another important parameter of MAX phases. $T_m$ can be calculated using elastic constants $C_{ij}$ applying the empirical formula [85]: $T_m = 354 + \frac{4.5(2C_{11}+C_{33})}{3}$. Fig. 6 (b) shows the $K_{min}$, $\gamma$ and $T_m$ as a function Ti contents. The minimum thermal conductivity is found to increase gradually where as $T_m$ is found to increase with Ti up to $x = 0.67$. The solid solution with $x = 0.67$ posses the highest melting temperature. The high values of Debye temperature and melting temperature implies that the solid solution at this particular composition posses a rather stiff lattice and a good thermal conductivity.

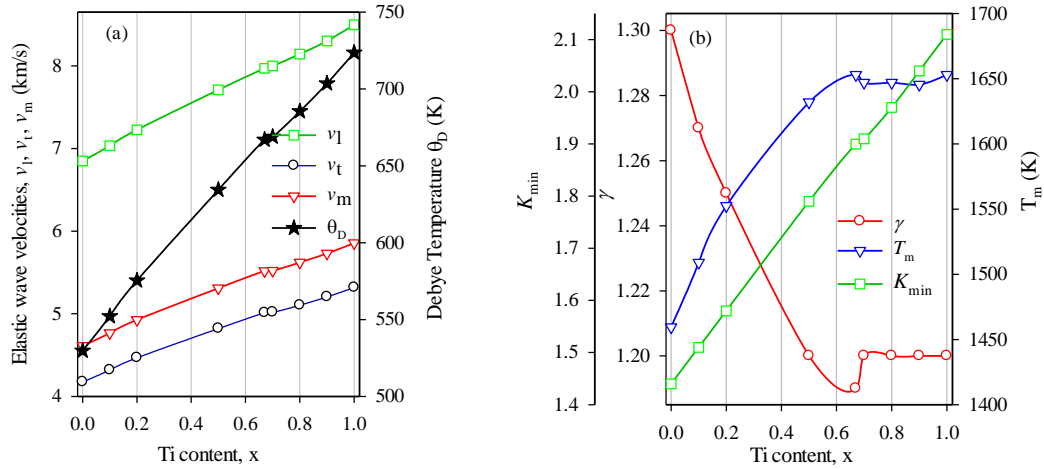

Fig. 6. Variation of (a) different sound velocities, Debye temperature and (b) minimum thermal conductivity, Grüneisen parameter, melting temperature with Ti contents ($x$).

3.5 *Charge transport properties*

The calculated electrical conductivity and Seebeck coefficient electronic specific heat, electronic part of thermal conductivity and ZT values of $(Zr_{1-x}Ti_x)_2AlC$ are presented in Fig. 7. The electrical conductivity of metals is usually calculated usingthe Boltzman theory, where the electrons are treated semi-classically and can be interpreted in terms of the average distance an electron traverses before it is scattered. A lot of research has been carried out about the saturation of electrical resistivity [86-90]. It is obvious from Fig. 7 (a) that the electrical conductivity of $(Zr_{1-x}Ti_x)_2AlC$ exhibit ordinary metallic behavior up to ~ 300 K and then the electrical conductivity is saturated. The sign of Seebeck coefficient is often used to qualitatively determine the sign of the dominant charge carriers. The Seebeck coefficient (Fig. 7 (b)) of $(Zr_{1-x}Ti_x)_2AlC$ (x = 1.0) is negative (n-type conduction) and almost constant up to ~ 400 K and then it become positive (p-type) and increase with temperature. On the other hand, the Seebeck coefficient for $(Zr_{1-x}Ti_x)_2AlC$ (x = 0.0, 0.5) are negative throughout whole temperature range studied and hence the charge carriers are electrons. The maximum Seebeck coefficient, -22 μV/K, is obtained at 400 K when 50% Zr is replaced by Ti. We achieved maximum power factor, $(S^2\sigma/\tau =11.1\times10^{10}$ Wm$^{-1}$K$^{-2}$s$^{-1})$ for ZrTiAlC at 400 K. The electronic heat capacity increases almost linearly with temperature for $(Zr_{1-x}Ti_x)_2AlC$ as illustrated in Fig. 7 (c). Thus the linear relation between the electronic heat capacity and temperature holds, i.e., $c_{el} = \gamma T$, where $\gamma$ is Sommerfeld constant [91, 92]. For $(Zr_{1-x}Ti_x)_2AlC$ compounds the electronic thermal conductivity increase with temperature and this reaffirms their metallic nature as shown in Fig. 7 (d). The values of electronic thermal conductivities for $Zr_2AlC$ and $Ti_2AlC$ at different temperature were very similar. The substitution of Zr by Ti is responsible for the reduction of thermal conductivity. We have also calculated the dimensionless figure of merit $(ZT = (S^2\sigma/\kappa_e)T)$ for the three compositions Zr$_2$AlC, Ti$_2$AlC and ZrTiAlC as shown in Fig. 7 (e). The obtained ZT values at 400 K for two end members Zr$_2$AlC and Ti$_2$AlC are 0.004×10$^{-2}$ and 3.9×10$^{-7}$, respectively whereas it is 2.0×10$^{-2}$ for ZrTiAlC. The ZT value of ZrTiAlC is small due to high thermal conductivity but 500 times higher than Zr$_2$AlC. Therefore, if the reduction of thermal conductivity by doping with suitable element is possible, ZrTiAlC could be a promising candidate as thermoelectric materials.

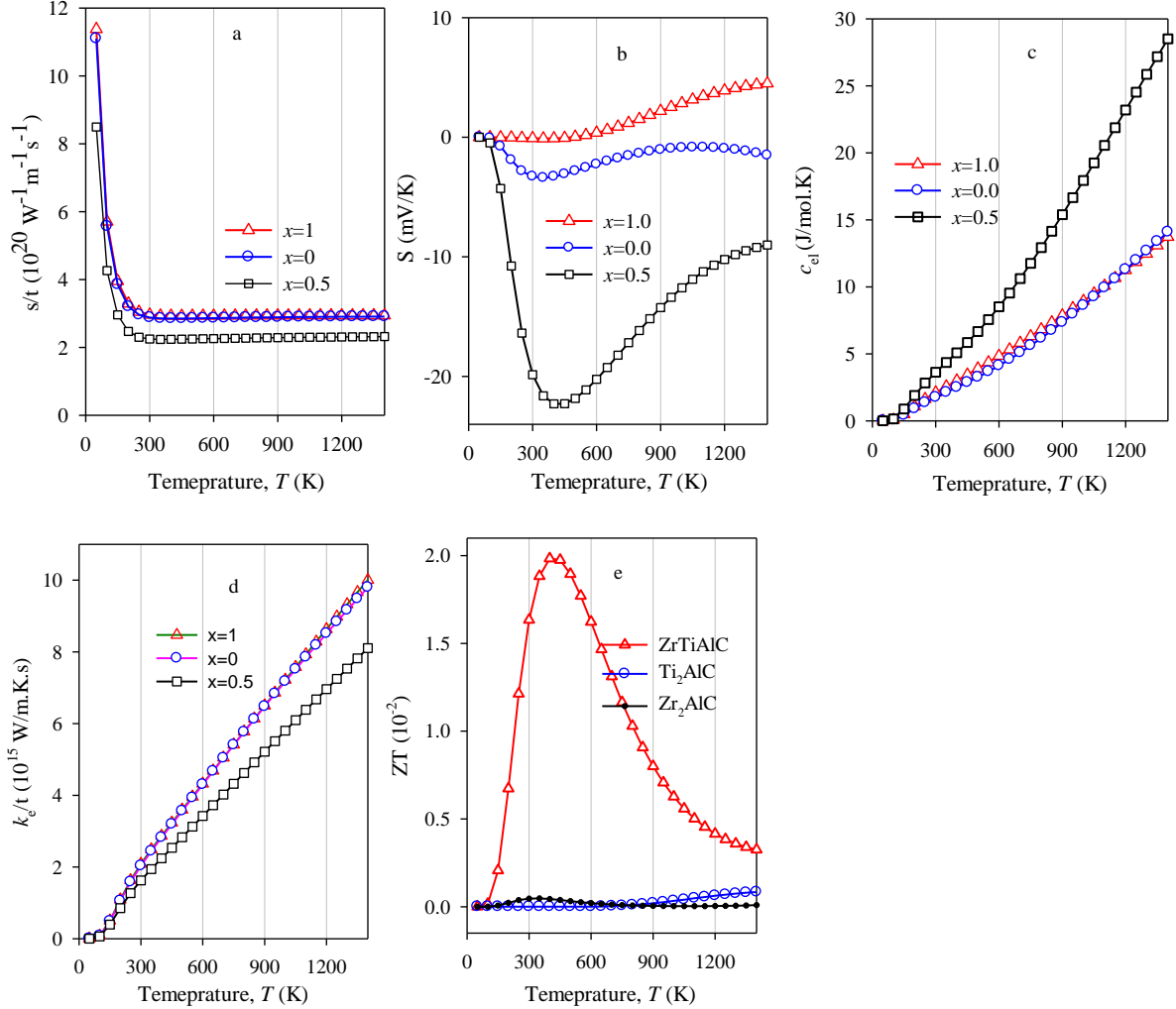

Fig. 7: Temperature dependence of (a) electrical conductivity and (b) Seebeck coefficient (c) electronic specific heat (d) electronic part of thermal conductivity and (e) ZT values for Zr$_{1-x}$Ti$_x$)$_2$AlC.

## 4. Conclusions

In this work, we have systematically studied the structural, elastic, electronic, thermodynamic and charge transport properties of recently synthesized (Zr$_{1-x}$Ti$_x$)$_2$AlC ($0 \leq x \leq 1$) solid solutions for the first time by first-principles DFT based calculations. The calculated lattice constants agree reasonably well with their experimental counterparts and decrease gradually with increasing Ti due mainly to smaller ionic radius. The stiffness constants and elastic parameters increase significantly in the Ti substituted solid solutions in comparison to Zr$_2$AlC. Particularly, the (Zr$_{0.33}$Ti$_{0.67}$)$_2$AlC compound is expected to possess the highest hardness compared to rest of the compounds. Elastic anisotropy factors are modified due to Ti substitutions. For instance, the anisotropic factor obtained from the compressibility shows both qualitative and quantitative variations with atomic substitution. The calculated DOS confirm the MAX compounds are metallic in nature with dominant contribution from the Zr/Ti-*d* orbitals. Electronic charge density mapping revealed the presence of mixed chemical bonding with covalent bonding between the Zr/Ti-d states and C-p states dominating. The strengthening of covalent bonding due to Ti substitution results improvement mechanical and thermal properties of $(Zr_{1-x}Ti_x)_2AlC$ solid. The Debye temperature $\Theta_D$ and minimum thermal conductivity $K_{min}$ of $(Zr_{1-x}Ti_x)_2AlC$ are found to increase as the Ti content increases, suggesting that the solid solutions posses a rather stiff lattice and good thermal conductivity. The obtained transport properties represent the metallic nature of $(Zr_{1-x}Ti_x)_2AlC$ compounds. The electrical resistivity was saturated after ~ 300 K. The Seebeck coefficient for $Zr_2AlC$ and ZrTiAlC were n-type for the whole temperature range studied whereas after ~ 400 K it was p-type for $Ti_2AlC$. The maximum Seebeck coefficient and hence maximum power factor was found for ZrTiAlC at 400 K and making it a promising candidate as thermoelectric material. Therefore, in this study we have found strong evidence that many important mechanical, thermal and transport features of $Zr_2AlC$ can be markedly improved by partially substituting Zr with Ti. We hope that our study will stimulates further theoretical and experimental studies on this solid solution system and these interesting materials will be tried for future industrial applications.